\documentclass[preprint,12pt]{elsarticle}

\usepackage{amssymb}
\usepackage{amsmath}
\usepackage{mathtools}
\usepackage{comment}
\usepackage[dvipsnames]{xcolor}

\newtheorem{theorem}{Theorem}

\journal{Ecological Complexity}

\begin{document}

\begin{frontmatter}



\title{Characterizing long transients in consumer-resource systems with group defense and discrete reproductive pulses}


\author{Jorge Arroyo-Esquivel$^{1}$, Alan Hastings$^2$, and Marissa L. Baskett$^2$}

\address{$^1$Department of Mathematics, University of California Davis}
\address{$^2$Department of Environmental Science and Policy, University of California Davis}

\begin{abstract}
During recent years, the study of long transients has been expanded in ecological theory to account for shifts in long-term behavior of ecological systems. These long transients may lead to regime shifts between alternative states that resemble the dynamics of alternative stable states for a prolonged period of time. One dynamic that potentially leads to long transients is the group defense of a resource in a consumer-resource interaction. Furthermore, time lags in the population caused by discrete reproductive pulses have the potential to produce long transients, either independently or in conjunction to the transients caused by the group defense. In this work, we analyze the potential for long transients in a model for a consumer-resource system in which the resource exhibits group defense and reproduces in discrete reproductive pulses. We develop this discrete-time model by discretizing a pulse differential equation. This system exhibits crawl-by transients near the extinction and carrying capacity states of resource. In addition, we identify a transcritical bifurcation in our system, under which a ghost limit cycle appears. These transients resemble stable states for a prolonged transient time period. We estimate the transient time of our system from these transients using perturbation theory. This work advances an understanding of how systems shift between alternate states and their duration of staying in a given regime and what ecological  dynamics may lead to long transients.
\end{abstract}




\end{frontmatter}


\section{Introduction}
One of the goals of mathematical modelling of ecological systems is to understand the fate or long term dynamics of such system. The main method to study such fate has been through the analysis of the attractors in a model \cite{ives_stability_2007}. Recent years have seen an increase in the interest of understanding non-attractor dynamics (hereafter transients) of the models, especially those that resemble an attractor for a long period of time (hereafter long transients) \cite{hastings_transient_2018}. Long transients have gained recognition as a theoretical tool to better describe population dynamics by allowing the study of dynamics that occur in a more biologically relevant timeframe \cite{morozov_long_2020}. In addition, an understanding of long transients can inform conservation and natural resource management goals. For example, identifying that a positively-valued long-term behavior observed in nature is actually a long transient and what causes it can guide management to prolong it \cite{francis_management_2021}.

Long transients often appear in the presence of a ``small'' (close to zero) parameter in the model \cite{morozov_long_2020}. One of the main challenges of identifying long transients is identifying such a small parameter, which may be a function of the biologically reasonable parameters, and thus may not be easily interpretable. For example, in ghost attractors, this small parameter is the difference between a bifurcation parameter and its bifurcation value \cite{morozov_long_2020}. While varying the parameter past such bifurcation leads to the destruction of an attractor, small differences the transient dynamics will resemble the attractor. In crawl-by attractors, the small parameter is determined by the degree to which the trajectory of the system is parallel to the stable manifold of a saddle node equilibrium at a given time \cite{morozov_long_2020}. In this case the system will behave similarly to such a stable manifold for a prolonged period of time before the unstable part of the trajectory leads to a change in the system behavior.

One behavior that has been demonstrated to lead to long transients in consumer-resource systems is group defense \cite{venturino_spatiotemporal_2013}. Group defense is a behavior where a resource population reduces the risk of individuals being predated by protecting each other. This behavior occurs in diverse animal taxa, which produce early-warning signals to detect predators, as is the case of colonial spiders \cite{uetz_antipredator_2002}, birds, \cite{robinson_coloniality_1985}, and mammals \cite{ebensperger_grouping_2002}. This behavior also occurs in producers such as kelp, where high densities of kelp lead to an increase in predators of kelp grazers, which induces cryptic behavior on such grazers and thus reduces grazing intensity \cite{karatayev_grazer_2021}.

Group defense transients might also depend on lags in population growth caused by discrete reproductive pulses. In some taxa that exhibit group defense, adult stages of the population may reproduce in discrete, seasonal pulses, such as is the case of kelp \cite{karatayev_grazer_2021} or bees \cite{kastberger_social_2008}. This can provide individuals to a population decades after stressful events which cause population declines, such as competitive exclusion of pioneer species in tropical rain forests \cite{dalling_longterm_2009}, or extreme weather events in phytoplankton \cite{ellegaard_long-term_2018}.

In this paper we characterize the long transients in a consumer-resource with both group defense and reproductive pulses. We first construct the model that describes a consumer-resource interaction where the resource exhibits group defense and has discrete reproductive pulses. Then, to illustrate the long transients present in this model, we identify a small parameter that describes each of the transients (crawl-by and ghost attractor), and we use this parameter to calculate the time the system remains in this long transient (hereafter transient time). Finding approximations for these parameters and transient times provides biological insight into how these long transients may arise in natural systems with the modelled dynamics. We conclude this paper with a discussion of these results and their biological implications.

\section{Model}
In this section we construct a consumer-resource model with group defense and discrete reproductive pulses. We previously explored a spatial, non-smooth version of this model to understand spread of kelp being grazed by urchins \cite{arroyo-esquivel_how_2021}. We consider the dynamics of adult consumer $P$ and adult resource $N$ densities through time. Adults of population $i=P,N$ experience a natural mortality at a rate $d_i$. In addition, consider that consumers consume resource following a unimodal Type IV Holling functional response that represents group defense with a decline in consumption at high resource densties \cite{andrews_mathematical_1968}. We let $\gamma_N$ be the attack rate of the consumer, and the maximum per-capita resource consumption occurs when $N=\frac{1}{\sigma_N}$.

Reproduction and recruitment of juvenile stages occur at discrete points in time. We model this recruitment as a pulse differential equation. Let $t=m$ be the periods at which the offspring recruit to the population. The number of consumer recruits is proportional to the amount of resource consumed at time $t=m$ with proportionality constant $\gamma_P$. Resource produce a per-capita number $R$ of recruits. We assume that $R>1-\exp(-d_N)$ in order to have a self-replenishing resource in the absence of consumers. For predation, a fraction of those offspring survive consumption with a probability following an exponential distribution with mean $\frac{1}{\gamma_S}$. Resource offspring also survive intracompetition from adults with carrying capacity proportional to $\frac{1}{\beta}$. 

Then, given $P_{m+1}^-$ as the density of consumers before the pulse and $P_{m+1}^+$ its density after the pulse (with analogous notation for resource, $N_{m+1}^-$ and $N_{m+1}^+$), the dynamics of the adult consumer and resource populations satisfy the following system of pulse differential equations:

\begin{equation}\label{eq:hybrid_model}
    \begin{split}
        \frac{d P}{d t}&=-d_PP,\\
        \frac{d N}{d t}&=-\frac{\gamma_N PN}{1+\sigma_N N^2}-d_NN,\\
        P^+_{m+1}&=P^-_{m+1}+\gamma_P\frac{P^-_{m+1}N^-_{m+1}}{1+\sigma_N N^{-2}_{m+1}},\\
        N^+_{m+1}&=N^-_{m+1}+R\frac{\exp\left(-\gamma_S P^-_{m+1}\right)}{1+\beta N^-_{m+1}}N^-_{m+1}.
    \end{split}
\end{equation}

We next transform Model \ref{eq:hybrid_model} into a discrete-time model. We can rewrite the continuous part of the Model \ref{eq:hybrid_model} as

\begin{equation}\label{eq:hybrid_model_continuous}
        \begin{split}
        \frac{1}{P}\frac{d P}{d t}&=-d_P,\\
        \frac{1}{N}\frac{d N}{d t}&=-\frac{\gamma_N P}{1+\sigma_N N^2}-d_N.
        \end{split}
\end{equation}

Following the derivation of \cite{cui_complex_2016}, we discretize the System \ref{eq:hybrid_model_continuous} as

\begin{equation}\label{eq:discretized_hybrid}
        \begin{split}
        P_{m+1}&=P_m\exp(-d_P),\\
        N_{m+1}&=N_m\exp(-d_N)\exp\left(-\frac{\gamma_N\exp(-d_P) P_m}{1+\sigma_N N_m^2}\right).
        \end{split}
\end{equation}

By taking $\delta_P=\exp(-d_P),\delta_N=\exp(-d_N)$ and using $P_{m+1}^-=P_m$ and $N_{m+1}^-=N_m$ as described in System \ref{eq:hybrid_model}, we arrive the following discrete-time model:

\begin{equation}\label{eq:spatial_Dimensional}
    \begin{split}
        P_{m+1}&=\delta_P P_m+\gamma_P\frac{P_{m}N_m}{1+\sigma_N N_{m}^2},\\
        N_{m+1}&=\delta_NN_m\exp\left(-\frac{\gamma_N\delta_P P_m}{1+\sigma_N N_m^2}\right)\\&+R\frac{\exp\left(-\gamma_S P_m\right)}{1+\beta N_m}N_m.
    \end{split}
\end{equation}

To simplify our analysis, we will study a nondimensional version of the model. For each $m$, let $p_m=\gamma_SP_m,n_m=\beta N_m$. Then, if $\gamma_p=\gamma_P/\beta,\gamma_n=\gamma_N\delta_P/\gamma_S,\sigma=\sigma_N/\beta^2$, our nondimensional version of the model is

\begin{equation}\label{eq:nonspatial_nonDimensional}
    \begin{split}
        p_{m+1}&=\delta_p p_m+\gamma_p\frac{p_{m}n_m}{1+\sigma n_{m}^2},\\
        n_{m+1}&=\delta_nn_m\exp\left(-\frac{\gamma_n p_m}{1+\sigma n_m^2}\right)+Rn_m\frac{\exp\left(- p_m\right)}{1+n_m}.
    \end{split}
\end{equation}

Note that we have also changed the indices of $\delta_i$ and $k_i$ in order to preserve clarity.

\section{Analysis and Results}
In this section we characterize the dynamics of Model \ref{eq:nonspatial_nonDimensional} and its potential for long transient dynamics. We identify two different classes of long transients, a crawl-by transient around the extinction of resource, and a ghost consumer-resource cycle.

Before we characterize these long transients, we first analyze the equilibria of the model. This model has up to four biologically relevant fixed points: a resource-only carrying capacity equilibrium $(0,n^*)$, an unstable extinction equilibrium $(0,0)$, and two possible unstable coexistence saddle equilibria $(p^{\vee\wedge},n^{\vee\wedge})$. We assume that the carrying capacity of resource is greater than the density at which consumption growth is its highest, i.e. $n^*>1/\sqrt{\sigma}$, such that group defense is relevant to resource populations below carrying capacity. Under this condition, the equilibria $(0,n^*)$ and $(p^\wedge,n^\wedge)$ go through a transcritical bifurcation at

\begin{equation}\label{eq:gamma_star}
    \gamma_p^*=(1-\delta_p)\frac{1+\sigma n^{*2}}{n^*}.
\end{equation}

In this case, the equilibria $(0,n^*)$ is stable for $\gamma_p<\gamma_p^*$ and unstable for $\gamma_p>\gamma_p^*$. See Appendix A for the expressions of these equilibria and their stability. This analysis allows us to better understand the nature of the transients we have identified.

\subsection{Crawl-by transients}

\begin{figure}\label{fig:long_transient}
    \centering
    \includegraphics[scale=0.6]{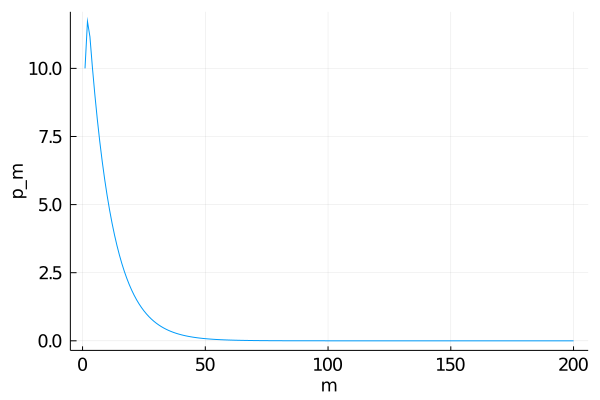}
    \includegraphics[scale=0.6]{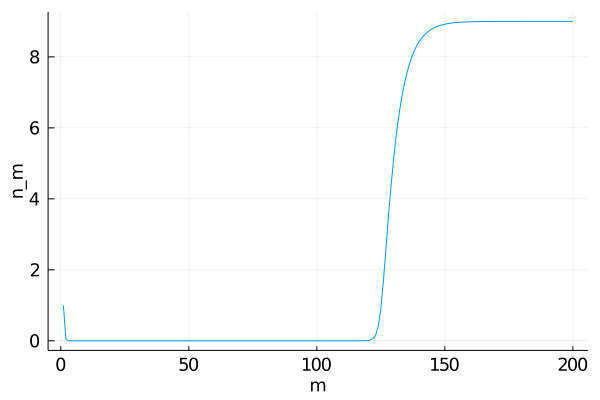}
    \caption{Time series of \textbf{a)} consumers $p_m$ and \textbf{b)} resource $n_m$ following System \ref{eq:nonspatial_nonDimensional} for 200 time steps ($m$). In this figure, $p_0=10,n_0=1,\delta_p=0.9,\gamma_p=1,\sigma=2.67,\delta_n=0.8,\gamma_n=1,R=2$.}
\end{figure}

Although the coextinction equilibrium is a saddle in the $n$-direction (which implies that $n$ will stay above 0), System \ref{eq:nonspatial_nonDimensional} can resemble a system where the resource is extinct for a long period of time when consumer density is high (Figure \ref{fig:long_transient}). This is an example of a long crawl-by transient. We determine how prevalent this behavior is in the following Theorem, proven in Appendix B.

\begin{theorem}\label{thm:transient_time}
Let $\varepsilon\ll1$. If $p_0$ is of order $\varepsilon^{-1}$ and $n_0$ of order $1$, then System \ref{eq:nonspatial_nonDimensional} goes through a crawl-by transient at the extinction of resource $n=0$. Recovery of resource will begin after a time of approximately \begin{equation}M=O\left(\frac{\log(\varepsilon)}{\log(\delta_p)}\right).\end{equation}
\end{theorem}

\begin{figure}\label{fig:recovery_vs_transient}
    \centering
    \includegraphics[scale=0.6]{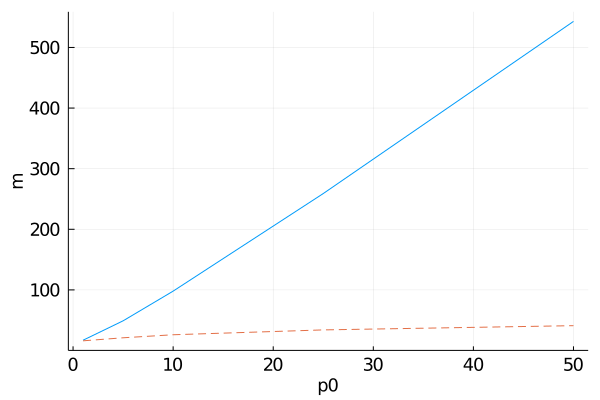}
    \caption{Resource recovery time (dashed line) and transient extinction time of resource (solid line) in System \ref{eq:nonspatial_nonDimensional} as a function of initial consumer density $p_0$, where transient extinction time of resource is defined as the time where resource is below 1\% of its carrying capacity. In this figure,  $\delta_p=0.9,\gamma_p=1,\sigma=2.67,\delta_n=0.8,\gamma_n=1,R=2$.}
\end{figure}

Although we are able to approximate the time where resource will recover, the transient ``extinction'' state can prolong itself for a longer time, as illustrated in Figure \ref{fig:long_transient}. We compare the time before resource recover and the transient extinction time (defined as the time before the resource recovers to above 1\% of its carrying capacity) for different resource densities in Figure 2. As $p_0$ increases, the gap between the period where resource recovery starts and where the transient extinction disappears increases, possibly because a bigger $p_0$ not only requires more time to decrease to a level where resource starts recovering, but also because resource levels will become increasingly smaller, causing recovery to be a slower process.

Another question of these transient dynamics is at what densities they occur. If we approximate the exponential consumption terms as

$$\exp\left(-\frac{\gamma_np}{1+\sigma n^2}\right)\sim 1-\frac{\gamma_np}{1+\sigma n^2},$$
$$\exp(-p)\sim 1-p,$$

\noindent we can approximate this critical density by finding the minimum value where these two expressions become nonpositive. For a given initial resource density $n_0$, this value is

\begin{equation}\label{eq:consumer_threshold}
    p_0=\overline{p}=\max\left(\frac{1+\sigma n_0^2}{\gamma_n},1\right).
\end{equation}

We call the value $\overline{p}$ the consumer threshold. Note that this analysis does not provide any information on the long term dynamics of the model. We will show in the following theorem that the long term dynamics seen in Figure \ref{fig:long_transient} did not depend on the initial conditions of the model.

\begin{theorem}\label{thm:global_attractor}
System \ref{eq:nonspatial_nonDimensional} has a compact, connected global attractor in the first quadrant $M=\{(p,n)\in\mathbb{R}^2:p\geq0,n\geq0\}$.
\end{theorem}

\begin{figure}\label{fig:nonlinear_attractor}
    \centering
    \includegraphics[scale=0.32]{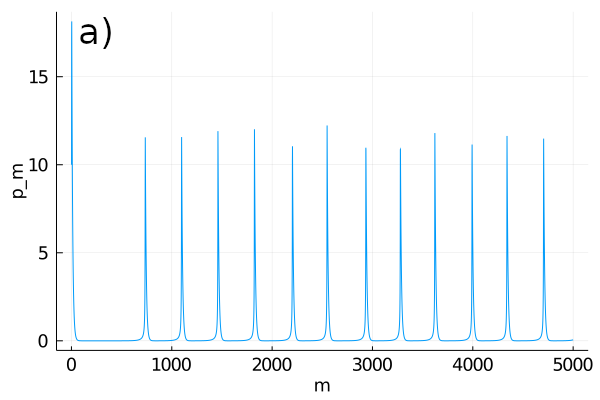}
    \includegraphics[scale=0.32]{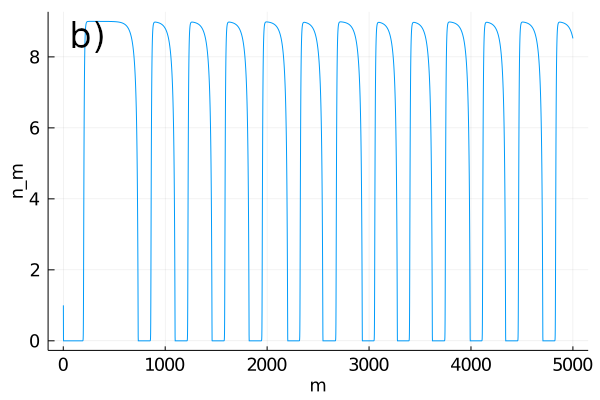}
    \includegraphics[scale=0.32]{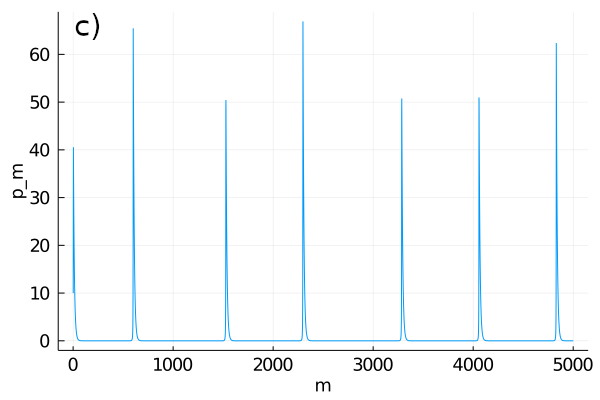}
    \includegraphics[scale=0.32]{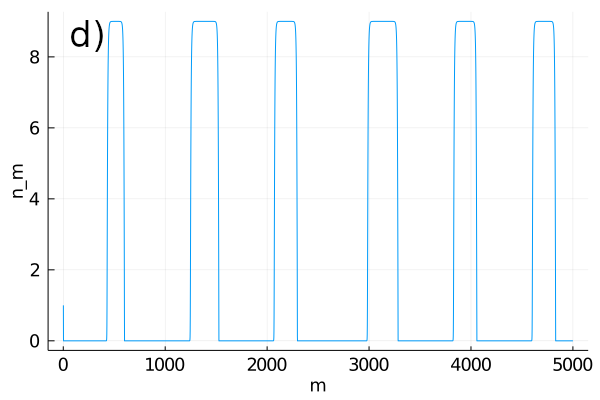}
    \caption{Time series of consumers $p_m$ (\textbf{a} and \textbf{c}) and resource $n_m$ (\textbf{b} and \textbf{d}) following System \ref{eq:nonspatial_nonDimensional} for 5000 time steps ($m$). In this figure, $p_0=10,n_0=1,\delta_p=0.9,\sigma=2.67,\delta_n=0.8,\gamma_n=1,R=2,$ and the consumer conversion intensity $\gamma_p=3$ (\textbf{a} and \textbf{b}) and $\gamma_p=8$ (\textbf{c} and \textbf{d}).}
\end{figure}

See Appendix C for the proof of this theorem. Theorem \ref{thm:global_attractor} implies that, when $\gamma_p<\gamma_p^*$, System \ref{eq:nonspatial_nonDimensional} will go towards carrying capacity of resource and extinction of consumers. On the other hand, when $\gamma_p>\gamma_p^*$, there are no stable fixed points in the first quadrant. Thus, Theorem \ref{thm:global_attractor} implies the existence of a nonlinear attractor, which we can describe based on numerical observations, as seen in Figure \ref{fig:nonlinear_attractor}.

When $\gamma_p>\gamma_p^*$, the resource population is able to reach a maximum density of carrying capacity and stay there for a prolonged period of time (Figure 3). However, after the consumer reaches a high enough density, the resource population collapses and passes through a transient extinction phase. This cycle repeats itself through time, but at each repetition, the amplitude of consumer density varies. We hypothetize that this variation in amplitude is caused by the system having a long periodicity. In addition, increasing $\gamma_p$ increases the period between each oscillation. This is consistent with the implication from Theorem \ref{thm:transient_time} that a higher consumer density causes the resource to stay around the extinction equilibrium for a longer period of time.

Figure 3 also shows that the system can stay around the resource-only equilibrium for a prolonged time. We approximate this time in the following Theorem, proven in Appendix D.

\begin{theorem}\label{thm:crawl_by_K}
Let $\gamma_p>\gamma_p^*$, where $\gamma_p^*$ is defined by Equation \ref{eq:gamma_star} and let 

\begin{equation}\lambda_1=\delta_p+\gamma_p\frac{n^*}{1+\sigma n^{*2}}.
\end{equation}

Then, if $(p_0,n_0)=(\varepsilon,n^*-\varepsilon)$ for $0<\varepsilon\ll1$, System \ref{eq:nonspatial_nonDimensional} goes a crawl-by transient at the resource-only equilibrium $n=n^*$. resource will start decaying after a time of approximately

\begin{equation}
    M=O\left(\frac{\log\left(\frac{1}{\varepsilon}\right)}{\log(\lambda_1)}\right).
\end{equation}
\end{theorem}

\subsection{Ghost attractors}
Theorem \ref{thm:global_attractor} ensures that, when $\gamma_p<\gamma_p^*$, System \ref{eq:nonspatial_nonDimensional} will converge to the stable equilibrium $(0,n^*)$. However, when $\gamma_p^*-\gamma_p\ll 1$, this convergence can take a significantly longer time, as can be seen in Figure 4. Before the system reaches the equilibrium, the dynamics resemble pseudo-oscillations similar to those observed in Figure \ref{fig:nonlinear_attractor} when $\gamma_p>\gamma_p^*$. 

\begin{figure}\label{fig:ghost_attractor}
    \centering
    \includegraphics[scale=0.6]{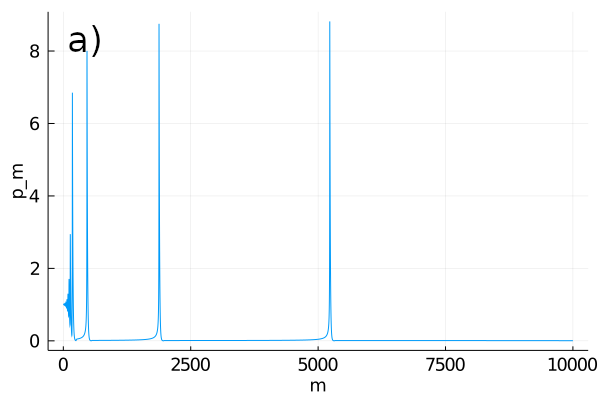}
    \includegraphics[scale=0.6]{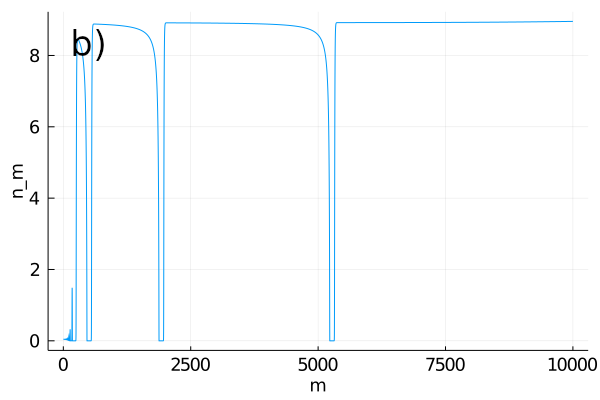}
    \caption{Time series of \textbf{a)} consumers $p_m$ and \textbf{b)} resource $n_m$ following System \ref{eq:nonspatial_nonDimensional} for 10000 time steps. In this figure we consider $p_0=10,n_0=1,\delta_p=0.9,\gamma_p=0.9912\gamma_p^*,\sigma=2.67,\delta_n=0.8,\gamma_n=1,R=2$. Although we know that the system will converge to the equilibrium point, this convergence takes over 5000 time steps.}
\end{figure}

Given limitations of available analytical tools for exact derivation of limit cycles in discrete-time models, we approximate the time spent in the ghost attractor $\tau$ by considering a power law for the time spent in a limit cycle \cite{medeiros_trapping_2017}:

\begin{equation}\label{eq:power_law}
    \tau(\gamma_p)= A(\gamma_p^*-\gamma_p)^{-B}.
\end{equation}

Whenever $n_M>n^\wedge$ and $p_M<p^\wedge$, System \ref{eq:nonspatial_nonDimensional} shows that $n_{k+1}>n_k$ and $p_{k+1}<p_k$ for all $k>M$. Therefore, we identify the time the system escapes the ghost attractor as $\tau=\min\{M:n_M>n^\wedge,p_M<p^\wedge\}$. Figure 5 shows that this approximation using the power law provides a reasonable approximation.

\begin{figure}\label{fig:epsilon_vs_tau}
    \centering
    \includegraphics[scale=0.6]{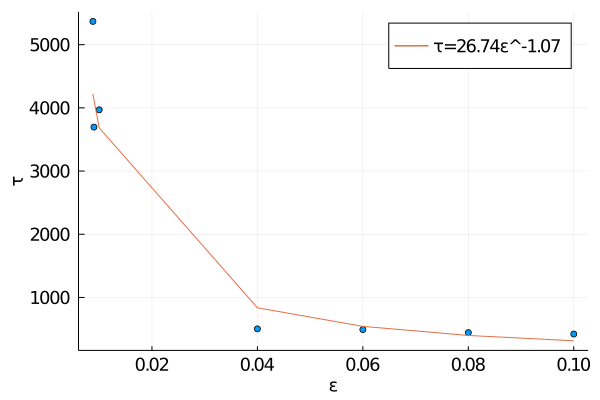}
    \caption{Approximation of transient time ($\tau$) following Equation \ref{eq:power_law}, where $\varepsilon=\gamma_p^*-\gamma_p$. In this figure, $p_0=10,n_0=1,\delta_p=0.9,\sigma=2.67,\delta_n=0.8,\gamma_n=1,R=2$.}
\end{figure}

\section{Discussion}
In this work we have identified two types of long transients, crawl-by transients and ghost attractors, that can appear in a consumer-resource system with group defense with discrete reproductive pulses. Our long-term dynamics are qualitatively different from those found in \cite{cui_complex_2016}, where they identified a variety of bifurcations and chaotic dynamics. The key differences between the two models are that, while that of \cite{cui_complex_2016} models reproduction as a continuous process and integrates the density-dependent growth for a case with overcompensation in discrete time, our model considers reproduction as a discrete event and has a saturating density-dependent function. The model in \cite{cui_complex_2016} is a discrete-time model similar to the Ricker model, where increased reproduction rates lead to unstable dynamics \cite{ricker_stock_1954}. When modelling reproduction as a discrete process with saturating (Beverton-Holt style) density dependence rather than overcompensation, our analysis did not suggest that increased reproduction numbers leads to instabilities in our model.

In addition, discrete reproduction events are one of the main reasons we see the long transients analyzed in this model. The crawl-by transient observed at high consumer densities (Theorem \ref{thm:transient_time}) is caused by a sudden crash in the adults of the resource population, which is followed by a slow crash of the consumer population due to its inability to find enough resource for self-replacement. Although the adult resource population is almost nonexistent, the few remaining individuals eventually lead to an increase the resource population when the consumer population becomes small enough.

The other reason long transients appear in this model is due to the self-replacement of consumers depending on the ability of resource to defend themselves. The Type IV Holling functional response produces a bifurcation on the proportionality constant $\gamma_p$ at the value $\gamma_p^*$ given by Equation \ref{eq:gamma_star}. This constant can be associated with the conversion capability of consumers, i.e.\ the amount of energy invested in reproduction activities. When the conversion capability of consumers is too small ($\gamma_p<\gamma_p^*$), group defense of resource will prevent self-replacement of consumers at high resource densities, which will lead to collapse of consumers. When this conversion capability is high enough, self-replacement can be satisfied, and the consumer-resource cycles of Figure 3 will occur. These cycles and their condition for existence resemble those found in other models where a mechanism of group defense of resource is considered \cite{ajraldi_modeling_2011,venturino_minimal_2011,venturino_spatiotemporal_2013}.

In the case where the system presents consumer-resource cycles, the resource-dominated phase will include a crawl-by transient when the conversion capability is close to this bifurcation value (Theorem \ref{thm:crawl_by_K}). This will follow by a crash of the resource population, where the consumer-dominated phase appears and presents the crawl-by transient previously described. This behavior presents an alternative perspective to the concept of alternate stable states \cite{beisner_alternative_2003}, where the different ``alternative stable states'' constitute long transients, which may resemble stable states during a long period of time, which then transition into the other phase and stay in a different long transient. In reality, stochasticity may render this juvenile survival when rare impossible or accelerate consumer death, which may lead the model to a stable state in a shorter period of time \cite{reimer_noise_2021}.

When the conversion capability of consumers is smaller than the critical value $\gamma_p^*$ but close to it, these quasiperiodic orbits do not disappear completely and stay as ghost attractors. This ghost attractor stays until the resource density surpasses a given threshold (the equilibrium value $n^\wedge$) and the system enters the basin of attraction of the resource-only equilibrium. The emergence of this ghost attractor is caused by group defense, because in its absence ($\sigma=0$), the unstable equilibria that cause the quasiperiodic orbits $(p^{\vee\wedge},n^{\vee\wedge})$ do not exist. In their absence, resource population density will consistently increase and the consumer density decrease.

The estimation of the transient time of the ghost attractor shows one of the limitations of our analysis, as the theory to study limit cycles in discrete-time systems is not developed enough to precisely analyze the transient time of this ghost attractor. Given the seasonality of the reproduction and recruitment for many organisms \cite{arreguin-sanchez_growth_1992,cameron_reproduction_1986,russell_seasonality_1977,wallace_seasonal_1985}, a continuous-time model may not properly reflect the biological dynamics we are interested in. Despite this challenge, the expression for the transient time found for transient limit cycles in continuous-time systems in \cite{medeiros_trapping_2017} is a reasonably accurate fit in our model. The transient times of the ghost attractor found in our work are similar to those found in the predator-prey model with group defense of \cite{venturino_spatiotemporal_2013}. However, our biological mechanisms differ, as their transients could be attributed to search time of prey from the predators through space, a feature not explicitly modelled in our work. In contrast, the length of the ghost attractor in our model can be attributed to the length of the crawl-by transients that are part of the cycle itself, which are periods of low population growth for either the consumer or the resource.

In conclusion, we show how long transients can appear in predator-prey systems with group defense and discrete recruitment pulses. A possible extension of this model is to explicitly consider the dynamics of the juvenile stages through a continuous-time model, which could give a more accurate approximation of the transient times found in this paper. A multi-stage model would also allow exploration of the effect of relative adult versus juvenile vulnerability to consumption on the transient dynamics.

\section*{Acknowledgements}
This research is supported under California Sea Grant \#R/HCE-15: ``A multi-pronged approach to kelp recovery along California’s north coast". Arroyo-Esquivel thanks the University of Costa Rica for their support through the development of this paper.




\begin{thebibliography}{}

\bibitem[Ajraldi et~al., 2011]{ajraldi_modeling_2011}
Ajraldi, V., Pittavino, M., and Venturino, E. (2011).
\newblock Modeling herd behavior in population systems.
\newblock {\em Nonlinear Analysis: Real World Applications}, 12(4):2319--2338.

\bibitem[Andrews, 1968]{andrews_mathematical_1968}
Andrews, J.~F. (1968).
\newblock A mathematical model for the continuous culture of microorganisms
  utilizing inhibitory substrates.
\newblock {\em Biotechnology and Bioengineering}, 10(6):707--723.
\newblock \_eprint:
  https://onlinelibrary.wiley.com/doi/pdf/10.1002/bit.260100602.

\bibitem[Arreguin-Sanchez, 1992]{arreguin-sanchez_growth_1992}
Arreguin-Sanchez, F. (1992).
\newblock Growth and seasonal recruitment of {Octopus} maya on {Campeche}
  {Bank}, {Mexico}.
\newblock {\em Naga, the ICLARM Quarterly}, 15(2):31--34.

\bibitem[Arroyo-Esquivel et~al., 2021]{arroyo-esquivel_how_2021}
Arroyo-Esquivel, J., Baskett, M.~L., McPherson, M., and Hastings, A. (2021).
\newblock How far to build it before they come? {Analyzing} the impact of the
  {Field} of {Dreams} hypothesis in bull kelp restoration.
\newblock {\em Under review}.

\bibitem[Beisner et~al., 2003]{beisner_alternative_2003}
Beisner, B.~E., Haydon, D.~T., and Cuddington, K. (2003).
\newblock Alternative stable states in ecology.
\newblock {\em Frontiers in Ecology and the Environment}, 1(7):376--382.
\newblock \_eprint:
  https://esajournals.onlinelibrary.wiley.com/doi/pdf/10.1890/1540-9295\%282003\%29001\%5B0376\%3AASSIE\%5D2.0.CO\%3B2.

\bibitem[Cameron, 1986]{cameron_reproduction_1986}
Cameron, R.~A. (1986).
\newblock Reproduction, larval occurrence and recruitment in {Caribbean} sea
  urchins.
\newblock {\em Bulletin of Marine Science}, 39(2):332--346.

\bibitem[Cui et~al., 2016]{cui_complex_2016}
Cui, Q., Zhang, Q., Qiu, Z., and Hu, Z. (2016).
\newblock Complex dynamics of a discrete-time predator-prey system with
  {Holling} {IV} functional response.
\newblock {\em Chaos, Solitons \& Fractals}, 87:158--171.

\bibitem[Dalling and Brown, 2009]{dalling_longterm_2009}
Dalling, J. and Brown, T. (2009).
\newblock Long‐{Term} {Persistence} of {Pioneer} {Species} in {Tropical}
  {Rain} {Forest} {Soil} {Seed} {Banks}.
\newblock {\em The American Naturalist}, 173(4):531--535.
\newblock Publisher: The University of Chicago Press.

\bibitem[Ebensperger and Wallem, 2002]{ebensperger_grouping_2002}
Ebensperger, L.~A. and Wallem, P.~K. (2002).
\newblock Grouping increases the ability of the social rodent, {Octodon} degus,
  to detect predators when using exposed microhabitats.
\newblock {\em Oikos}, 98(3):491--497.
\newblock \_eprint:
  https://onlinelibrary.wiley.com/doi/pdf/10.1034/j.1600-0706.2002.980313.x.

\bibitem[Ellegaard and Ribeiro, 2018]{ellegaard_long-term_2018}
Ellegaard, M. and Ribeiro, S. (2018).
\newblock The long-term persistence of phytoplankton resting stages in aquatic
  ‘seed banks’: {Persistence} of phytoplankton resting stages.
\newblock {\em Biological Reviews}, 93(1):166--183.

\bibitem[Francis et~al., 2021]{francis_management_2021}
Francis, T.~B., Abbott, K.~C., Cuddington, K., Gellner, G., Hastings, A., Lai,
  Y.-C., Morozov, A., Petrovskii, S., and Zeeman, M.~L. (2021).
\newblock Management implications of long transients in ecological systems.
\newblock {\em Nature Ecology \& Evolution}, 5(3):285--294.
\newblock Number: 3 Publisher: Nature Publishing Group.

\bibitem[Gobbino and Sardella, 1997]{gobbino_connectedness_1997}
Gobbino, M. and Sardella, M. (1997).
\newblock On the {Connectedness} of {Attractors} for {Dynamical} {Systems}.
\newblock {\em Journal of Differential Equations}, 133(1):1--14.

\bibitem[Hastings et~al., 2018]{hastings_transient_2018}
Hastings, A., Abbott, K.~C., Cuddington, K., Francis, T., Gellner, G., Lai,
  Y.-C., Morozov, A., Petrovskii, S., Scranton, K., and Zeeman, M.~L. (2018).
\newblock Transient phenomena in ecology.
\newblock {\em Science}, 361(6406).

\bibitem[Ives and Carpenter, 2007]{ives_stability_2007}
Ives, A.~R. and Carpenter, S.~R. (2007).
\newblock Stability and {Diversity} of {Ecosystems}.
\newblock {\em Science}, 317(5834):58--62.
\newblock Publisher: American Association for the Advancement of Science
  Section: Review.

\bibitem[Karatayev et~al., 2021]{karatayev_grazer_2021}
Karatayev, V.~A., Baskett, M.~L., Kushner, D.~J., Shears, N.~T., Caselle,
  J.~E., and Boettiger, C. (2021).
\newblock Grazer behavior can regulate large-scale patterns of community
  states.
\newblock {\em Ecology Letters}, In press.

\bibitem[Kastberger et~al., 2008]{kastberger_social_2008}
Kastberger, G., Schmelzer, E., and Kranner, I. (2008).
\newblock Social {Waves} in {Giant} {Honeybees} {Repel} {Hornets}.
\newblock {\em PLOS ONE}, 3(9):e3141.
\newblock Publisher: Public Library of Science.

\bibitem[Khan et~al., 2016]{khan_bifurcations_2016}
Khan, A.~Q., Ma, J., and Xiao, D. (2016).
\newblock Bifurcations of a two-dimensional discrete time plant-herbivore
  system.
\newblock {\em Communications in Nonlinear Science and Numerical Simulation},
  39:185--198.

\bibitem[Magal and Zhao, 2005]{magal_global_2005}
Magal, P. and Zhao, X.-Q. (2005).
\newblock Global {Attractors} and {Steady} {States} for {Uniformly}
  {Persistent} {Dynamical} {Systems}.
\newblock {\em SIAM Journal on Mathematical Analysis}, 37(1):251--275.

\bibitem[Medeiros et~al., 2017]{medeiros_trapping_2017}
Medeiros, E.~S., Caldas, I.~L., Baptista, M.~S., and Feudel, U. (2017).
\newblock Trapping {Phenomenon} {Attenuates} the {Consequences} of {Tipping}
  {Points} for {Limit} {Cycles}.
\newblock {\em Scientific Reports}, 7(1):42351.
\newblock Number: 1 Publisher: Nature Publishing Group.

\bibitem[Morozov et~al., 2020]{morozov_long_2020}
Morozov, A., Abbott, K., Cuddington, K., Francis, T., Gellner, G., Hastings,
  A., Lai, Y.-C., Petrovskii, S., Scranton, K., and Zeeman, M.~L. (2020).
\newblock Long transients in ecology: {Theory} and applications.
\newblock {\em Physics of Life Reviews}, 32:1--40.

\bibitem[Murakami, 2007]{murakami_stability_2007}
Murakami, K. (2007).
\newblock Stability and bifurcation in a discrete-time predator–prey model.
\newblock {\em Journal of Difference Equations and Applications},
  13(10):911--925.
\newblock Publisher: Taylor \& Francis \_eprint:
  https://doi.org/10.1080/10236190701365888.

\bibitem[Reimer et~al., 2021]{reimer_noise_2021}
Reimer, J.~R., Arroyo-Esquivel, J., Jiang, J., Scharf, H.~R., Wolkovich, E.~M.,
  Zhu, K., and Boettiger, C. (2021).
\newblock Noise can create or erase long transient dynamics.
\newblock {\em Theoretical Ecology}.

\bibitem[Ricker, 1954]{ricker_stock_1954}
Ricker, W.~E. (1954).
\newblock Stock and {Recruitment}.
\newblock {\em Journal of the Fisheries Research Board of Canada},
  11(5):559--623.

\bibitem[Robinson, 1985]{robinson_coloniality_1985}
Robinson, S.~K. (1985).
\newblock Coloniality in the {Yellow}-{Rumped} {Cacique} as a {Defense} against
  {Nest} {Predators}.
\newblock {\em The Auk}, 102(3):506--519.

\bibitem[Russell et~al., 1977]{russell_seasonality_1977}
Russell, B.~C., Anderson, G. R.~V., and Talbot, F.~H. (1977).
\newblock Seasonality and recruitment of coral reef fishes.
\newblock {\em Marine and Freshwater Research}, 28(4):521--528.

\bibitem[Uetz et~al., 2002]{uetz_antipredator_2002}
Uetz, G.~W., Boyle, J., Hieber, C.~S., and Wilcox, R.~S. (2002).
\newblock Antipredator benefits of group living in colonial web-building
  spiders: the ‘early warning’ effect.
\newblock {\em Animal Behaviour}, 63(3):445--452.

\bibitem[Venturino, 2011]{venturino_minimal_2011}
Venturino, E. (2011).
\newblock A minimal model for ecoepidemics with group defense.
\newblock {\em Journal of Biological Systems}, 19(04):763--785.
\newblock Publisher: World Scientific Publishing Co.

\bibitem[Venturino and Petrovskii, 2013]{venturino_spatiotemporal_2013}
Venturino, E. and Petrovskii, S. (2013).
\newblock Spatiotemporal behavior of a prey–predator system with a group
  defense for prey.
\newblock {\em Ecological Complexity}, 14:37--47.

\bibitem[Wallace, 1985]{wallace_seasonal_1985}
Wallace, C.~C. (1985).
\newblock Seasonal peaks and annual fluctuations in recruitment of juvenile
  scleractinian corals.
\newblock {\em Marine Ecology Progress Series}, 21(3):289--298.

\end{thebibliography}


\appendix
\section{Fixed points of System \ref{eq:nonspatial_nonDimensional} and their stability}
The fixed points of System \ref{eq:nonspatial_nonDimensional} $(p,n)$ satisfy the equations

$$p=\delta_pp+\gamma_p\frac{pn}{1+\sigma n^2}$$
$$n=\delta_n n\exp\left(-\frac{\gamma_np}{1+\sigma n^2}\right)+Rn\frac{\exp(-p)}{1+n}.$$

If $p=0$, then the second equation gives us two solutions for $n$, $n=0$ and

\begin{equation}\label{eq:n_star}n^*=\frac{R}{1-\delta_n}-1.\end{equation}

If $p\neq 0$, then the first equation has two solutions for $n$ given by

\begin{equation}\label{eq:n_pm}n^{\vee\wedge}=\frac{\gamma_p}{2(1-\delta_p)\sigma}\left(1\pm\sqrt{1-\frac{4\sigma(1-\delta_p)^2}{\gamma_p^2}}\right)\end{equation}

\noindent where $n^\vee$ corresponds to the solution with a $-$ sign and $n^\wedge$ to the solution with a $+$ sign. These solutions are positive whenever $\gamma_p\geq 2\sqrt{\sigma}(1-\delta_p)$. In such case, plugging $n^{\vee\wedge}$ into the second equation provides us with the following expression:

$$\delta_n\exp\left(-\frac{\gamma_n p}{1+\sigma n^{\pm2}}\right)+\frac{R}{1+n^{\vee\wedge}}\exp(-p)=1.$$

Then there is an unique value $p^{\vee\wedge}$ that solves the trascendental equation

\begin{equation}\label{eq:p_pm}p^{\vee\wedge}=\log\left(\frac{R}{(1+n^{\vee\wedge})\left(1-\delta_n\exp\left(-\frac{\gamma_n p^{\vee\wedge}}{1+\sigma n^{\pm2}}\right)\right)}\right).\end{equation}

This equation in $p$ has an unique solution as the function \begin{equation}\label{eq:fp}f(p)=p-\log\left(\frac{R}{(1+n^{\vee\wedge})\left(1-\delta_n\exp\left(-\frac{\gamma_n p}{1+\sigma n^{\pm2}}\right)\right)}\right)\end{equation} \noindent is monotonic for $p$ and satisfies $\lim_{p\rightarrow{-\infty}}f(p)<0$ and $\lim_{p\rightarrow{\infty}}f(p)>0$. For it to be biologically relevant, we also require $\lim_{p\rightarrow{0}}f(p)<0$, which will occur when

\begin{equation}R>(1-\delta_n)(1+n^{\vee\wedge})\end{equation}

\noindent or, after reorganizing the terms, $n^{\vee\wedge}<n^*$.

The Jacobian of the system $J$ is the following:

\begin{equation}\label{eq:Jacobian}J(p,n)=\left(\begin{array}{cc}
     \delta_p+\gamma_p\frac{n}{1+\sigma n^2}&\gamma_pp\frac{1-\sigma n^2}{(1+\sigma n^2)^2}  \\
   -\frac{\delta_n\gamma_nn}{1+\sigma n^2} \exp\left(-\frac{\gamma_np}{1+\sigma n^2}\right)-\frac{Rn\exp(-p)}{1+n}& \delta_n\exp\left(-\frac{\gamma_np}{1+\sigma n^2}\right)\left(1+\frac{2\sigma\gamma_n pn^2}{(1+\sigma n^2)^2}\right)+\frac{R\exp(-p)}{(1+n)^2}
\end{array}\right).\end{equation}

From here, the extinction equilibrium satisfies

\begin{equation}J(0,0)=\left(\begin{array}{cc}
     \delta_p &  0\\
     0 & \delta_n+R
\end{array}\right)\end{equation}

\noindent which has eigenvalues $\delta_p<1$ and $\delta_n+R>1$. Therefore the extinction equilibrium is a saddle. For the resource-only equilibrium, the upper right term of the Jacobian matrix equals 0 whenever $p=0$. Therefore $J(0,n^*)$ is a triangular matrix, with the eigenvalues being the diagonal terms

\begin{equation}\lambda_1=\delta_p+\gamma_p\frac{n^*}{1+\sigma n^{*2}},\end{equation}
\begin{equation}\lambda_2=\delta_n+\frac{(1-\delta_n)^2}{R}.\end{equation}

Because $R>1-\delta_n$, $0<\lambda_2<1$. $\lambda_1$, on the other hand, $\lambda_1$ will produce a change in stability when

\begin{equation}\gamma_p=\gamma_p^*=(1-\delta_p)\frac{1+\sigma n^{*2}}{n^{*}}.\end{equation}

In this case, the equilibrium is stable whenever $\gamma_p<\gamma_p^*$ and a saddle when $\gamma_p>\gamma_p^*$. Because $n^*>1/\sqrt{2\sigma}$, plugging $\gamma_p=\gamma_p^*$ in Equation \ref{eq:n_pm}, we have that $n^*=n^\wedge$. Based on the conditions for $(p^\wedge,n^\wedge)$ to be biologically reasonable, this implies that at $\gamma_p=\gamma_p^*$, the system goes through a transcritical bifurcation, where the carrying capacity $(0,n^*)$ changes stability.

The trascendental equation that describes $p^{\vee\wedge}$ renders it impossible to analyze them directly. However, a numerical exploration in Figure A.6 shows that these equilibria are unstable for $\gamma_p<\gamma_p^*$ and $(p^\wedge,n^\wedge)$ becomes stable for  $\gamma_p>\gamma_p^*$.

This transcritical bifurcation occurs with almost any combination of parameters in our region of interest. To show this, we perform a similar analysis as those in \cite{khan_bifurcations_2016,murakami_stability_2007}. When $\gamma_p=\gamma_p^*$, we can rewrite our system in diagonal form and centered around the origin by making the change of variables:

\begin{equation}x_m=\frac{\lambda_2-1}{n^*\left(1-\delta_n+\frac{\delta_n\gamma_n}{1+\sigma n^{*2}}\right)}p_m\end{equation}
\begin{equation}y_m=p_m+n_m-n^*\end{equation}

\noindent provided that $\left(1-\delta_n+\frac{\delta_n\gamma_n}{1+\sigma n^{*2}}\right)\neq 0$. Otherwise, we let $x_m=p_m,y_m=n_m-n^*$. In both cases, this lets use write our System \ref{eq:nonspatial_nonDimensional} as

\begin{equation}\left(\begin{array}{c}
     x_{m+1}  \\
     y_{m+1} 
\end{array}\right)=\left(\begin{array}{cc}
    1 & 0 \\
    0 & \lambda_2
\end{array}\right)\left(\begin{array}{c}
     x_{m}  \\
     y_{m} 
\end{array}\right)+\hbox{h.o.t}.\end{equation}

We can expand this system to include the parameter as a dynamical factor with eigenvalue 1 as $\mu_m\equiv\gamma_p-\gamma_p^*$. The central limit theorem gives us that $y_m=h(x_m,\mu_m)$ for some function $h=O((x_m+\mu_m)^2)$. Because $x_m$ is a multiple of $p_m$, its dynamics follow the same trend, and can be approximated up to $O((x_m+\mu_m)^3)$ as:

\begin{equation}x_{m+1}=f(x_m,\mu_m)=x_m+\frac{\gamma_p^*(1-\sigma n^{*2})}{(1+\sigma n^{*2})^2}x_m^2+\frac{\sigma n^{*2}(n^*-\gamma_p^*)}{(1+n^{*2})^2}x\mu_m+O((x_m+\mu_m)^3)\end{equation}

Equation A.14 satisfies that $f_x(0,0)=1,f_\mu(0,0)=0,f_{xx}(0,0)\neq0$ Because we assume that $n^*>1/\sqrt{\sigma}$, and $f_{x\mu}(0,0)\neq0$ except when $n^*=\gamma_p^*$. Plugging this value Equation \ref{eq:gamma_star}, we get that the condition $n^*=\gamma_p^*$ holds only when $n^*= \sqrt{\frac{1-\delta_p}{\sigma\delta_p}}$. 

Therefore, whenever $n^*\neq \sqrt{\frac{1-\delta_p}{\sigma\delta_p}}$, the system goes through a transcritical bifurcation between $(0,n^*)$ and $(p^\wedge,n^\wedge)$ as $\gamma_p$ passes through $\gamma_p^*.$

\begin{figure}\label{fig:bifurcation_diagram}
    \centering
    \includegraphics[scale=0.4]{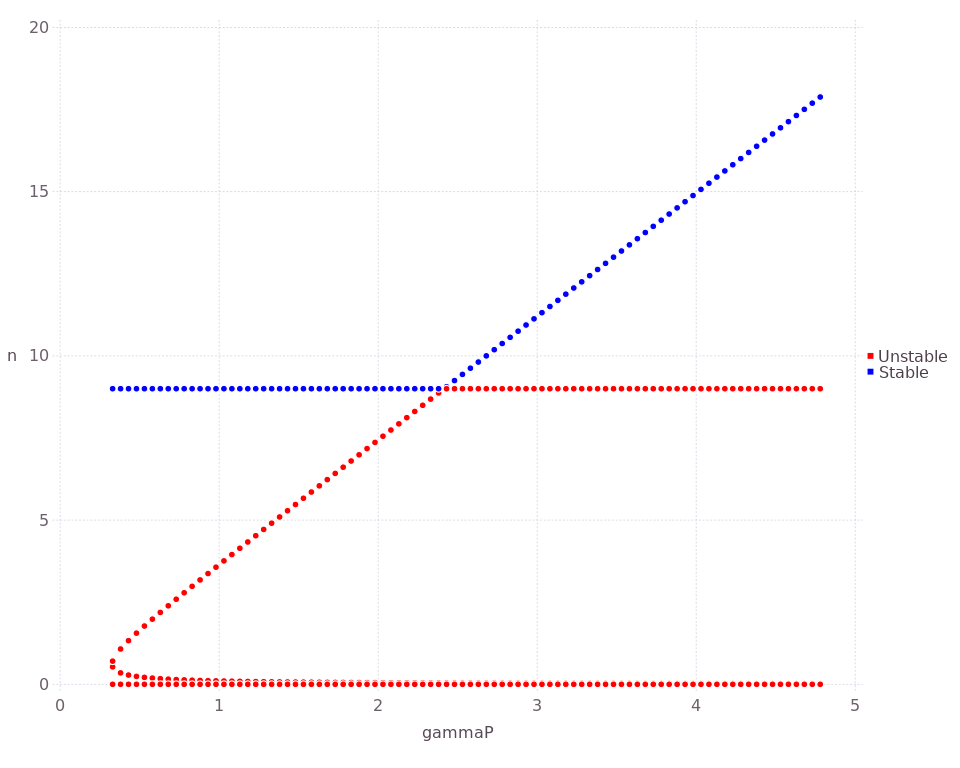}
    \caption{Numerical values of different equilibria for $n$ and their stability as we vary $\gamma_p$ in our region of interest. The red points correspond to unstable equilibria, whereas the blue points correspond to stable equilibria. In this figure, $\delta_p=0.9,\sigma=2.67,\delta_n=0.8,\gamma_n=1,R=2$.}
\end{figure}

\section{Proof of Theorem \ref{thm:transient_time}}
If $p_0=O(\varepsilon^{-1})$, plugging $O(\varepsilon^{-1})$ into the formula for $n_1$ gives us that $n_1=o(\varepsilon)$. Plugging $o(\varepsilon)$ into the formula for $p_2$ gives us that:

\begin{equation}p_2=\delta_pp_1+o(\varepsilon).\end{equation}

In addition, if $n_m=o(\varepsilon)$, the equation for $n_{m+1}$ satisfies:

\begin{equation}\frac{n_{m+1}}{n_m}=O\left(\delta_n\exp\left(-\gamma_n p_m\right)+R\exp(-p_m)\right).\end{equation}

While $p_m=O(\varepsilon^{-1})$, this expression will satisfy $n_{m+1}/n_m<1$. We can thus assume that the expression

\begin{equation}p_{m+1}=\delta_p p_m+o(\varepsilon)\end{equation}

\noindent is satisfied until $p_{m+1}=O(1)$. Therefore, when $p_m=O(\varepsilon^{-1})$, the consumer population time evolution can be approximately solved as

\begin{equation}p_m=\frac{\delta_p^m}{\varepsilon}+o(\varepsilon).\end{equation}

This expression stops working when $p_m=O(1)$, and thus resource will start a recovery afterwards. We can estimate the order of magnitude of such $m$ by plugging $p_m=1$ above. Solving for $m$ gives us that

\begin{equation}m=\frac{\log(\varepsilon)}{\log(\delta_p)}\end{equation}

\noindent which is the expression that proves the theorem.

\section{Proof of Theorem \ref{thm:global_attractor}}
To show the existence of this theorem, we use Theorem 2.9 of \cite{magal_global_2005}. To do this, we consider the first quadrant $M$ as a metric subspace of $\mathbb{R}^2$ with metric $d(x,y)=\|x-y\|_2$ induced by the Euclidean norm. Let $T:M\rightarrow M$ be given by

\begin{equation}\label{eq:T}T\left(\begin{array}{c}
     p\\
     n
\end{array}\right)=\left(\begin{array}{c}
     p\left(\delta_p+\frac{\gamma_p n}{1+\sigma n^2}\right)\\
     n\left(\delta_n\exp\left(-\frac{\gamma_np}{1+\sigma n^2}\right)+\frac{R\exp(-p)}{1+n}\right)
\end{array}\right).\end{equation}

We show that $T$ is a point dissipative, compact map on $M$. Because $M$ is a subspace of $\mathbb{R}^2$, compactedness is trivial. A map is point dissipative if there is a bounded set $B_0\subset M$ such that $B_0$ attracts each point in $M$. To show $T$ is point dissipative, we find such bounded set $B_0$.

Let \begin{equation}\label{eq:growth_factors}\begin{split}f_p(n)&=\delta_p+\frac{\gamma_p n}{1+\sigma n^2},\\f_n(p,n)&=\delta_n\exp\left(-\frac{\gamma_np}{1+\sigma n^2}\right)+\frac{R\exp(-p)}{1+n}.\end{split}\end{equation}

Note that $f_n<1$ whenever $n>n^*$, where $n^*$ is given by Equation \ref{eq:n_star}. This implies that $n$ is attracted by the set $[0,n^*]$. Without loss of generality, we assume that $n\in[0,n^*]$. Suppose that $p>p^*$, where $p^*$ is

\begin{equation}\label{eq:huge_p}
    p^*=\max\left(\frac{1+\sigma n^{*2}}{\gamma_n},1\right)=\frac{1}{\varepsilon}.
\end{equation}

A similar argument to that of the proof for Theorem \ref{thm:transient_time} shows that in this case, the consumer population will satisfy $p=O(1)$ in time $\ln(\varepsilon)/ln(\delta_p)$. In particular, $p<p^*$ after a period of time. In addition, if $p<p^*$ but $Tp>p^*$, $f_p$ satisfies:

\begin{equation}f_p(n)\leq\delta_p+\frac{\gamma_p}{2\sqrt{\sigma}}\end{equation}

\noindent for any $n$. This implies that $Tp\leq \left(\delta_p+\frac{\gamma_p}{2\sqrt{\sigma}}\right)p^*$. Let $\tau$ be the period of time such that $T^\tau\left(\delta_p+\frac{\gamma_p}{2\sqrt{\sigma}}\right)p^*<p^*$. Therefore $p$ is attracted by the set $[0,(\delta_p+\frac{\gamma_p}{2\sqrt{\sigma}})^\tau p^*]$. Then, the bounded rectangle 

\begin{equation}\label{eq:B0}
    B_0:=\left[0,\left(\delta_p+\frac{\gamma_p}{2\sqrt{\sigma}}\right)^\tau p^*\right]\times[0,n^*]
\end{equation}

\noindent is an attracting set in $M$. Therefore, $T$ is a point dissipative map.

Therefore, Theorem 2.9 of \cite{magal_global_2005} implies that there is a compact global attractor in $M$. Finally, Because $M$ is locally connected, Theorem 4.5 of \cite{gobbino_connectedness_1997} implies that the global attractor is connected, which completes the proof.

\section{Proof of Theorem \ref{thm:crawl_by_K}}
Let $x_m=p_m,y_m=n^*-n_m$. Because $\|(x_m,y_m)\|=O(\varepsilon)$, System \ref{eq:nonspatial_nonDimensional} can be approximated by the linearized system:

\begin{equation}\left(\begin{array}{c}
     x_{m+1}  \\
     y_{m+1} 
\end{array}\right)\sim J(0,n^*)\left(\begin{array}{c}
     x_{m}  \\
     y_{m} 
\end{array}\right),\end{equation}

where the Jacobian $J(0,n^*)$ is described by Equation \ref{eq:Jacobian}. The calculations of Appendix A show that the Jacobian $J(0,n^*)$ is a lower triangular matrix, with eigenvalues

\begin{equation}\lambda_1=\delta_p+\gamma\frac{n^*}{1+\sigma n^{*2}}.\end{equation}
\begin{equation}\lambda_2=\delta_n+\frac{(1-\delta_n)^2}{R}\end{equation}

\noindent and eigenvectors

\begin{equation}v_1=\left(\begin{array}{c}
     u  \\
     1 
\end{array}\right),v_2=\left(\begin{array}{c}
     0  \\
     1 
\end{array}\right)\end{equation}

\noindent where

\begin{equation}
    u=\frac{\lambda_2-\lambda_1}{n^*\left(1-\delta_n+\frac{\delta_n\gamma_n}{1+\sigma n^{*2}}\right)}.
\end{equation}

This system has for solution the expression

\begin{equation}\label{eq:linearized_solution}\left(\begin{array}{c}
     x_m  \\
     y_m
\end{array}\right)=a\lambda_1^mv_1+b\lambda_2^mv_2,\end{equation}

\noindent where $a,b$ are constants. If we let $m=0$, then we can solve the linear system

\begin{equation}\left(\begin{array}{c}
     \varepsilon  \\
     \varepsilon
\end{array}\right)=\left(\begin{array}{cc}
     u&0  \\
     1&1 
\end{array}\right)\left(\begin{array}{c}
     a  \\
     b
\end{array}\right),\end{equation}

\noindent which has solutions

\begin{equation}\label{eq:initial_conditions_constants}
    \left(\begin{array}{c}
     a  \\
     b
\end{array}\right)=\left(\begin{array}{cc}
     \frac{1}{u}&0  \\
     -\frac{1}{u}&1 
\end{array}\right)\left(\begin{array}{c}
     \varepsilon  \\
     \varepsilon
\end{array}\right).\end{equation}

In particular, this gives us that $a=\frac{\varepsilon}{u}$. Because $\gamma_p>\gamma_p^*$, then $\lambda_1>1$, and $\lambda_2<1$. Therefore, for big $m$, System \ref{eq:linearized_solution} can be approximated as

\begin{equation}\left(\begin{array}{c}
     x_m  \\
     y_m
\end{array}\right)\sim a\lambda_1^mv_1=\frac{\varepsilon \lambda_1^m}{u}\left(\begin{array}{c}
     u  \\
     1
\end{array}\right).\end{equation}

Thus, System \ref{eq:nonspatial_nonDimensional} will stay near the resource-only equilibrium as long as $\|(x_m,y_m)\|=O(\varepsilon)$. In particular, the System will escape the saddle point when $x_m=O(1)$. Plugging in $x_m=1$ into the approximated solution lets us find $M$ that solves the equation

\begin{equation}1=\varepsilon\lambda_1^M.\end{equation}

This has for solution

\begin{equation}M=\frac{\log\left(\frac{1}{\varepsilon}\right)}{\log(\lambda_1)}\end{equation}

\noindent which is the expression that proves the theorem.
\end{document}